\documentclass[a4paper,12pt]{article}
\pdfoutput=1
\usepackage{graphics,psfrag}
\usepackage{amsmath,amsthm,amssymb,epsfig,euscript,array,cite,cancel,color}
\usepackage{cases,empheq}
\usepackage[colorlinks=true]{hyperref}
\usepackage{verbatim}
\usepackage{ulem}
\usepackage[vcentermath]{youngtab}   

\usepackage[usenames,dvipsnames]{xcolor}  

\usepackage[paper=a4paper,dvips,top=2.54cm,left=2.74cm,right=2.34cm,
    foot=1cm,bottom=2.54cm]{geometry}

\theoremstyle{definition} 

\theoremstyle{definition}

\numberwithin{trial}{subsection}

\numberwithin{dtl}{subsection}

\theoremstyle{remark}

\newcounter{multieqs}

\newcommand{\be}{\begin{equation}}
\newcommand{\ee}{\end{equation}}
\newcommand{\eq}[1]{(\ref{#1})}
\newcommand{\bit}{\begin{itemize}}  \newcommand{\eit}{\end{itemize}}

\newcommand{\ket}[1]{|#1 \rangle}

\newcommand{\bm}[1]{\mbox{\boldmath $#1$}}
\newcommand{\rf}[1]{(\ref{#1})}

\def\bd{\begin{document}}
\def\ed{\end{document}}
\def\nn{\nonumber}
\def\bea{\begin{eqnarray}}
\def\eea{\end{eqnarray}}
\let\bm=\bibitem

\def\la{\langle}
\def\ra{\rangle}

\def\npb#1#2#3{Nucl. Phys. {\bf{B#1}} #3 (#2)}
\def\plb#1#2#3{Phys. Lett. {\bf{#1B}} #3 (#2)}
\def\prl#1#2#3{Phys. Rev. Lett. {\bf{#1}} #3 (#2)}
\def\prd#1#2#3{Phys. Rev. {D \bf{#1}} #3 (#2)}
\def\cmp#1#2#3{Comm. Math. Phys. {\bf{#1}} #3 (#2)}
\def\cqg#1#2#3{Class. Quantum Grav. {\bf{#1}} #3 (#2)}
\def\nppsa#1#2#3{Nucl. Phys. B (Proc. Suppl.) {\bf{#1A}}#3 (#2)}
\def\ap#1#2#3{Ann. of Phys. {\bf{#1}} #3 (#2)}
\def\ijmp#1#2#3{Int. J. Mod. Phys. {\bf{A#1}} #3 (#2)}
\def\rmp#1#2#3{Rev. Mod. Phys. {\bf{#1}} #3 (#2)}
\def\mpla#1#2#3{Mod. Phys. Lett. {\bf A#1} #3 (#2)}
\def\jhep#1#2#3{J. High Energy Phys. {\bf #1} #3 (#2)}
\def\atmp#1#2#3{Adv. Theor. Math. Phys. {\bf #1} #3 (#2)}

\def\N{{\cal N}}
\def\sst{\scriptscriptstyle}
\def\thetabar{\bar\theta}
\def\Tr{{\rm Tr}}
\def\one{\mbox{1 \kern-.59em {\rm l}}}

\def\a{\alpha}      \def\da{{\dot\alpha}}  \def\dA{{\dot A}}
\def\b{\beta}       \def\db{{\dot\beta}}
\def\g{\gamma}  \def\G{\Gamma}  \def\dc{{\dot\gamma}}
\def\d{\delta}  \def\D{\Delta}  \def\ddt{\dot\delta}
\def\e{\epsilon}        \def\ve{\varepsilon}
\def\f{\phi}    \def\F{\Phi}    \def\vvf{\f}
\def\h{\eta}
\def\k{\kappa}
\def\l{{\lambda}} \def\L{\Lambda}
\def\m{\mu} \def\n{\nu}
\def\om{\omega}
\def\p{\pi} \def\P{\Pi}
\def\r{\rho}
\def\s{\sigma}  \def\S{\Sigma}
\def\t{\tau}
\def\th{\theta} \def\Th{\Theta} \def\vth{\vartheta}
\def\X{\Xeta}
\def\z{\zeta}

\def\na{\nabla}

\def\cA{{\cal A}} \def\cB{{\cal B}} \def\cC{{\cal C}}
\def\cD{{\cal D}} \def\cE{{\cal E}} \def\cF{{\cal F}}
\def\cG{{\cal G}} \def\cH{{\cal H}} \def\cI{{\cal I}}
\def\cJ{{\cal J}} \def\cK{{\cal K}} \def\cL{{\cal L}}
\def\cM{{\cal M}} \def\cN{{\cal N}} \def\cO{{\cal O}}
\def\cP{{\cal P}} \def\cQ{{\cal Q}} \def\cR{{\cal R}}
\def\cS{{\cal S}} \def\cT{{\cal T}} \def\cU{{\cal U}}
\def\cV{{\cal V}} \def\cW{{\cal W}} \def\cX{{\cal X}}
\def\cY{{\cal Y}} \def\cZ{{\cal Z}}

\def\ua{{\underline{\alpha}}}
 \def\ub{\underline{\phantom{\alpha}}\!\!\!\beta}
\def\uc{\underline{\phantom{\alpha}}\!\!\!\gamma}
\def\um {{\underline{\mu}}}
\def\ud{{\underline{\delta}}}
\def\ue{\underline\epsilon}
\def\una{{\underline a}}\def\uA{{\underline A}}
\def\unb{{\underline b}}\def\uB{{\underline B}}
\def\unc{{\underline c}}\def\uC{{\underline C}}
\def\und{{\underline d}}\def\uD{{\underline D}}
\def\une{{\underline e}}\def\uE{{\underline E}}
\def\unf{{\underline{\phantom{e}}\!\!\!\! f}}\def\uF{{\underline F}}
\def\unm{{\underline m}\def\uM{\underline M}}
\def\unn{{\underline n}\def\uN{\underline N}}
\def\unp{{\underline{\phantom{a}}\!\!\! p}}\def\uP{{\underline P}}
\def\unq{{\underline{\phantom{a}}\!\!\! q}}
\def\uQ{{\underline{\phantom{A}}\!\!\!\! Q}}
\def\uH{{\underline{H}}}
\def\uM{{\underline{M}}}
\def\uN{{\underline{N}}}

\def\As {{A \hspace{-6.4pt} \slash}\;}
\def\bs {{b \hspace{-6.4pt} \slash}\;}
\def\Ds {{D \hspace{-6.4pt} \slash}\;}
\def\Gts {{\Gt \hspace{-6.4pt} \slash}\;}
\def\ds {{\del \hspace{-6.4pt} \slash}\;}
\def\ss {{\s \hspace{-6.4pt} \slash}\;}
\def\ks {{ k \hspace{-6.4pt} \slash}\;}
\def\ps {{p \hspace{-6.4pt} \slash}\;}
\def\xs {{x \hspace{-6.4pt} \slash}\;}
\def\pas {{{p_1} \hspace{-6.4pt} \slash}\;}
\def\pbs {{{p_2} \hspace{-6.4pt} \slash}\;}
\def\cFs {{{\cal F} \hspace{-6.4pt} \slash}\;}

\def\Ah{{\hat{A}}}
\def\Dh{{\hat{D}}}
\def\Gh{{\hat{G}}}
\def\Fh{{\hat{F}}}
\def\Ih{{\hat{I}}}
\def\Jh{{\hat{J}}}
\def\Kh{{\hat{K}}}
\def\Lh{{\hat{L}}}
\def\Ph{{\hat{P}}}
\def\Rh{{\hat{R}}}
\def\Vh{{\hat{V}}}
\def\Xh{{\hat{X}}}

\def\ah{{\hat{a}}}
\def\bh{{\hat{b}}}
\def\ch{{\hat{c}}}
\def\gh{{\hat{g}}}
\def\dh{{\hat{d}}}
\def\hh{{\hat{h}}}
\def\uh{{\hat{u}}}
\def\vh{{\hat{v}}}
\def\xh{{\hat{x}}}
\def\yh{{\hat{y}}}
\def\zh{{\hat{z}}}
\def\ph{{\hat{p}}}
\def\qh{{\hat{q}}}
\def\thh{{\hat{t}}}
\def\xih{\hat{\xi}}
\def\Psih{\hat{\Psi}}
\def\mh{{\hat{m}}}
\def\nh{{\hat{n}}}
\def\ih{{\hat{i}}}
\def\jh{{\hat{j}}}
\def\kh{{\hat{k}}}
\def\aah{{\hat{\alpha}}}
\def\bbh{{\hat{\beta}}}
\def\ggh{{\hat{\gamma}}}
\def\llh{{\hat{\ell}}}
\def\ph{{\hat{p}}}

\def\psit{\tilde{\psi}}
\def\Psit{\tilde{\Psi}}
\def\Psibt{\tilde{\bar{Psi}}}

\def\st{\tilde{\sigma}}

\def\delt{\tilde{\delta}}
\def\Phit{\tilde{\Phi}}
\def\Phitb{\overline{\tilde{Phi}}}
\def\tht{\tilde{\th}}
\def\lt{\tilde{\l}}
\def\chit{\tilde{\chi}}
\def\phit{\tilde{\phi}}

\def\At{\tilde{A}}
\def\Bt{\tilde{B}}
\def\Ct{\tilde{C}}
\def\Dt{\tilde{D}}
\def\Et{\tilde{E}}
\def\Ft{\tilde{F}}
\def\Gt{\tilde{G}}
\def\Ht{\tilde{H}}
\def\It{\tilde{I}}
\def\Jt{\tilde{J}}
\def\Qt{\tilde{Q}}
\def\Rt{\tilde{R}}
\def\Mt{\tilde{M }}
\def\Nt{\tilde{N}}
\def\St{\tilde{S}}
\def\Vt{\tilde{V}}
\def\Xt{\tilde{X}}
\def\at{\tilde{a}}
\def\ct{\tilde{c}}
\def\dt{\tilde{d}}
\def\htt{\tilde{h}}
\def\ft{\tilde{f}}
\def\gt{\tilde{g}}
\def\pt{\tilde{p}}
\def\qt{\tilde{q}}
\def\vt{\tilde{v}}
\def\nt{\tilde{n}}
\def\ut{\tilde{u}}
\def\wt{\tilde{w}}
\def\zt{\tilde{z}}
\def\xt{\tilde{x}}
\def\yt{\tilde{y}}
\def\Psit{\tilde{\Psi}}
\def\vphit{\tilde{\varphi}}

\def\gamt{\tilde{\gamma}}

\def\Tt{\tilde{T}}

\def\eb{\bar{\epsilon}}
\def\delb{\bar{\partial}}
\def\thb{\bar{\theta}}
\def\Thb{{\bar{\Theta}}}
\def\mub{\bar{\mu}}
\def\lamb{\bar{\l}}
\def\psib{\bar{\psi}}
\def\sb{\bar{\sigma}}
\def\xib{\bar{\xi}}
\def\chib{\bar{\chi}}

\def\Psib{\bar{\Psi}}
\def\Phib{\bar{\Phi}}
\def\Lamb{\bar{\Lambda}}
\def\Sb{{\overline \Sigma}}
\def\cb{\bar{c}}
\def\hb{\bar{h}}
\def\qb{\bar{q}}
\def\wb{\bar{w}}
\def\zb{{\bar{z}}}
\def\Hb{\bar{H}}
\def\Qb{{\bar Q}}
\def\Omegab{\overline{\Omega}}
\def\ob{\overline{\omega}}
\def\Gab{{\bar{\Gamma}}}

\def\Ab{{\overline A}} \def\Bb{{\overline B}} \def\Cb{{\overline C}}
\def\Db{{\overline D}} \def\Eb{{\overline E}} \def\Fb{{\overline F}}
\def\Gb{{\overline G}}
\def\Ib{{\overline I}}
\def\Jb{{\overline J}} \def\Kb{{\overline K}} \def\Lb{{\overline L}}
\def\Mb{{\overline M}} \def\Nb{{\overline N}} \def\Ob{{\overline O}}
\def\Pb{{\overline P}}  \def\Rb{{\overline R}}
 \def\Tb{{\overline T}} \def\Ub{{\overline U}}
\def\Vb{{\overline V}} \def\Wb{{\overline W}} \def\Xb{{\overline X}}
\def\Yb{{\overline Y}} \def\Zb{{\overline Z}}

\def\fb{{\overline f}}
\def\gb{{\overline g}}
\def\mb{{\overline m}}
\def\lb{{\overline l}}
\def\yb{{\overline y}}

\def\ldel{{\overleftarrow{\del}}}
\def\rdel{{\overrightarrow{\del}}}
\def\ldeldel{{\overleftarrow{\del^2}}}
\def\rdeldel{{\overrightarrow{\del^2}}}
\def\ldelb{{\overleftarrow{\bar{\del}}}}
\def\rdelb{{\overrightarrow{\bar{\del}}}}

\def\ba{{\bf a}}
\def\bk{{\bf k}}
\def\bl{{\bf l}}
\def\bp{{\bf p}}
\def\bq{{\bf q}}
\def\br{{\bf r}}
\def\bt{{\bf t}}
\def\bu{{\bf u}}
\def\bv{{\bf v}}
\def\bx{{\bf x}}
\def\by{{\bf y}}
\def\bR{{\bf R}}
\def\bV{{\bf V}}

\def\bone{{\bf 1}}

\def\va{{\vec a}}
\def\vk{{\vec k}}
\def\vp{{\vec p}}
\def\vq{{\vec q}}
\def\vx{{\vec x}}
\def\vy{{\vec y}}
\def\vu{{\vec u}}
\def\vv{{\vec v}}

\def\vs{{\vec \sigma}}
\def\vtau{{\vec \tau}}

\newcommand{\ov}[1]{\overrightarrow{#1}}

\def\frA{\mathfrak{A}}
\def\frB{\mathfrak{B}}
\def\frC{\mathfrak{C}}
\def\frD{\mathfrak{D}}
\def\frE{\mathfrak{E}}
\def\frF{\mathfrak{F}}
\def\frG{\mathfrak{G}}
\def\frH{\mathfrak{H}}
\def\frM{\mathfrak{M}}
\def\frN{\mathfrak{N}}
\def\frR{\mathfrak{R}}
\def\frW{\mathfrak{W}}

\def\fra{\mathfrak{a}}
\def\frb{\mathfrak{b}}
\def\frf{\mathfrak{f}}
\def\frg{\mathfrak{g}}
\def\frh{\mathfrak{h}}
\def\frl{\mathfrak{l}}
\def\frs{\mathfrak{s}}
\def\fri{\mathfrak{i}}
\def\frj{\mathfrak{j}}

\def\ma{\mathfrak{a}}
\def\mg{\mathfrak{g}}
\def\mR{\mathfrak{R}}
\def\mN{\mathfrak{N}}

\def\d{\delta}\def\D{\Delta}\def\ddt{\dot\delta}

\def\pa{\partial} \def\del{\partial}
\def\xx{\times}
\def\uno{\mbox{1 \kern-.59em {\rm l}}}

\def\trp{^{\top}}
\def\inv{^{-1}}
\def\dag{{^{\dagger}}}
\def\pr{^{\prime}}

\def\rar{\rightarrow}
\def\lar{\leftarrow}
\def\lrar{\leftrightarrow}

\newcommand{\0}{\,\!}      
\def\one{1\!\!1\,\,}
\def\im{\imath}
\def\jm{\jmath}

\newcommand{\tr}{\mbox{tr}}
\newcommand{\slsh}[1]{/ \!\!\!\! #1}

\def\vac{|0\rangle}
\def\lvac{\langle 0|}

\def\hlf{\frac{1}{2}}
\def\ove#1{\frac{1}{#1}}

\def\Box{\square}
\def\CC {\mathbb{C}}
\def\FF {\mathbb{F}}
\def\RR{\mathbb{R}}
\def\NN{\mathbb{N}}
\def\ZZ{\mathbb{Z}}
\def\bb#1{{\bf #1}}
\def\bcomment#1{}
\def\bfhat#1{{\bf \hat{#1}}}
\def\VEV#1{\left\langle #1\right\rangle}

\newcommand{\ex}[1]{{\rm e}^{#1}} \def\ii{{\rm i}}

\newcommand{\lrbrk}[1]{\left(#1\right)}
\newcommand{\lrsbrk}[1]{\left[#1\right]}
\newcommand{\lrcbrk}[1]{\left\{#1\right\}}
\newcommand{\sfrac}[2]{{\textstyle\frac{#1}{#2}}}

\def\stw{{\sqrt{2}}}

\def\rf {{\rm f}}
\def\ri {{\rm i}}
\def\rj {{\rm j}}
\def\rk {{\rm k}}
\def\rl {{\rm l}}
\def\rs {{\scriptscriptstyle \rm S}}
\def\rt {{\scriptscriptstyle \rm T}}

\def\rQ {{\scriptscriptstyle \rm \cQ}}
\def\rR {{\scriptscriptstyle \rm \cR}}

\def\cQb{{\cal \Qb}}
\def\cRb{{\cal \Rb}}
\def\cWb{{\cal \Wb}}

\def\fd {{\rm N}}
\def\afd {{\overline{\rm N}}}

\def \II {I\hspace{-.1em}I\hspace{.1em}}
\def \IIA {\mbox{\II A\hspace{.2em}}}
\def \IIB {\mbox{\II B\hspace{.2em}}}
\def \gs {g^s}
\def \ls {\lambda^s}

\def \I {{\cal I}}
\def \qs {q\hspace{-.53em}/\hspace{.15em}}
\def \ks {k\hspace{-.53em}/\hspace{.15em}}
\def \YM {{\mbox{\tiny YM}}}
\def \gym {g_{\YM}}

\def \Lc {\L_c}
\def\IR{\relax{\rm I\kern-.18em R}}
\def \id {{\bf 1}}

\def\cci{\ell}
\def\ccj{\ell'}

\def \thbb{\overline{\th\th}}
\newcommand \ol{\overline}
\def \lamb{\bar{\lambda}}
\def \vphi{\varphi}
\def \lambh{\hat{\bar{\lambda}}}
\def \lh{\hat{\lambda}}
\def \dd{\ddagger}

\def \ad {\dot{a}}
\def \bd {\dot{b}}
\def \cd {\dot{c}}
\def  \ddd {\dot{d}}
\def \ed {\dot{e}}
\def \fd {\dot{f}}
\def \Bh {\hat{B}}
\def \zm {{(0)}}
\def \nz {{(\text{KK})}}
\def \3{{(3)}}
\def \diag {\text{diag}}
\def \inm {{(m^{-1})}}
\def \3{{(3)}}
\def \6{{(6)}}
\def \2{{(2)}}
\def \7{{(7)}}
\def \4{{(4)}}
\def\1{{(1)}}
\def\5{{(5)}}
\def\0{{(0)}}

\def\eh{{\hat{e}}}
\def\fh{{\hat{f}}}
\def\lh{{\hat{l}}}
\def\rh{{\hat{r}}}
\def\wh{{\hat{w}}}
\renewcommand{\mh}{{\hat{m}}}

\def \DBI{{\text{DBI}}}
\def\et{{\tilde{\e}}}
\def\w{{\wedge}}
\def\bbV{{\mathbb{V}}}
\def\M{{(\text{M})}}
\def\T{{(\text{T})}}

\def\Hbt{{\tilde{\bar{H}}}}
\def\Fbt{{\tilde{\bar{F}}}}

\def\fR{{\mathfrak{R}}}
\def\fg{{\mathfrak{g}}}
\def \sk {\textsc{k}}
\def\S{{\Sigma}}
\def\nb{{\nabla}}
\def\bB{{\bf B}}

\colorlet{1}{red}
\colorlet{2}{green}
\colorlet{3}{blue}
\colorlet{4}{cyan}
\def\bJ{{\bold J}}
\def\tR{{\text{(R)}}}
\def\bQ{{\bold Q}}
\def\bo{{\pmb{\omega}}}
\def\ADM{{\text{ADM}}}
\def\st{{\tilde\sigma}}
\def\Fbh{{\Phi_{\text{bh}}}}
\def\bC{{\bold C}}
\def\bep{{\pmb\epsilon}}
\def\EM{{(\text{EM})}}
\def \bTh {\bold \Theta}
\def\bL{{\bold L}}
\def\tG{{\tilde{\Gamma}}}
\def \vL {\mathfrak{L}}
\def\bE{{\bold E}}

\author{Wu-Zhong Guo\footnote{wzguo@cts.nthu.edu.tw}$~^a$ and
 Feng-Li Lin\footnote{fengli.lin@gmail.com; on leave from NTNU}$~^{b,c}$
\\
\\
{\small  $^a$ \it Physics Division, National Center for Theoretical Sciences,}
\\
{\small\it National Tsing-Hua University, Hsinchu 30013, Taiwan}
\\
\\
{\small $^b$ \it Department of Physics, National Taiwan Normal University, Taipei, 116, Taiwan}
\\
{\small $^c$ \it Department of Physics, California Institute of Technology, Pasadena, California 91125, USA}
}

\title{\bf Quantum measurement in two-dimensional conformal field theories: Application to quantum energy teleportation}

\date{}
\begin{document}
\maketitle

\abstract{ We construct a set of quasi-local measurement operators in 2D CFT, and then use them to proceed the quantum energy teleportation (QET) protocol and show it is viable. These measurement operators are constructed out of the projectors constructed from shadow operators, but further acting on   the product of two spatially separated primary fields. They are equivalently the OPE blocks in the large central charge limit up to some UV-cutoff dependent normalization but the associated probabilities of outcomes are UV-cutoff independent. We then adopt these quantum measurement operators to show that the QET protocol is viable in general. We also check the CHSH inequality a la OPE blocks.}

\thispagestyle{empty}
\newpage
\tableofcontents

\setcounter{equation}0

\section{Introduction}

     Quantum entanglement has been studied intensively in the past few years in quantum field theory (QFT) and many-body systems, partly inspired by the Ryu-Takayanagi formula of the holographic entanglement entropy \cite{Ryu:2006bv,Ryu:2006ef}, partly inspired by the new quantum order in the many-body condensed matter systems \cite{spt,Li}, and moreover by the connection of these twos \cite{MERA,Swingle:2012wq}. There are usually two ways to characterize the quantum entanglement. One is to evaluate the entanglement entropy or R\'enyi entropies of the reduced density matrix of a quantum state.   The other way is to treat the entanglement of quantum state as the resources for some quantum information tasks, which will help to enhance the efficiency of the similar tasks in the classical computation and communication, and to reduce the complexity. There are many classic examples in the earlier development of quantum information sciences, such as quantum teleportation \cite{Bennett:1992tv}, dense coding \cite{Bennett:1992dc} and so on.  However, most of these examples are performed for the few-qubit systems, and seldom for the QFT or many-body systems.

      In this work, we would like to explore the possibility of defining the quantum measurement operators in one of the special QFTs, i.e., the conformal field theory (CFT), so that one can generalize the quantum protocols for qubit systems to the ones in CFT. Some earlier effort along this direction can be found in \cite{Laiho:2000mc} for free QFT. Here our CFT is in general interacting theory and can be seen as the critical phases of many-body systems. Thus, our scheme can be thought as a precursor to perform the quantum information tasks in critical systems. For simplicity, we will apply our quantum measurement operators in CFT to one particular quantum information task, the so-called quantum energy teleportation (QET) \cite{Hotta:2008uk,Hotta:2009fz,Hotta:2009ppa,Hotta:2011xj}, for which Alice will send the energy (not the quantum state) to Bob by LOCC.   Note that a holographic version for holographic CFT has been studied in  \cite{Giataganas:2016bml} based on the   the so-called surface/state correspondence \cite{Miyaji:2015yva,Miyaji:2015fia}, which states that each (space-like) hypersurface in AdS space corresponds to a quantum state in the dual CFT.

      We propose that the OPE blocks formulated in \cite{opeblock} can be used as a set of local quantum measurements in the weak sense, i.e., just holds for ground state but not in the operator sense. The OPE blocks can be shown to be equivalent to be the projector operators $P_k$'s with $k$ labelling the outcomes, which are constructed in the shadow formalism \cite{shadows}, acting on the product of two separated local primary operators, i.e., $O_i(x_1)O_j(x_2)$.   The projectors $P_k$'s are not local but smear over the entire spacetime. However, in 2D CFTs they can be reduced to quasi-local ones over the causal diamond subtended by the interval $[x_1,x_2]$. The reason of the weak sense is that the set of OPE blocks cannot be complete. This is easy to see by the fact that the set of projectors constructed by shadow formalism is by construction complete, but the associated OPE blocks cannot be. Despite that, this is good enough to adopt them for the QET protocol by either in the weak sense or adopting the view of acting $P_k$'s on the excited state $O_i(x_1)O_j(x_2)|0\rangle$ initially prepared by Alice for QET, where $|0\rangle$ is CFT's ground state.

      As an application of these OPE block quantum measurement, we adopt them to proceed the QET protocols in 2D CFTs. We find that such QET task is viable. This encourage the experimental realization of QET in 2D critical systems such as the edge states of Quantum Hall Effect. Moreover, we also use these measurement operators to show that one cannot violate CHSH inequality.

     In the following the paper is organized as follows. In section \ref{section2} we will review the issues of POVM in QFT, and then propose the OPE blocks as the set of quasi-local quantum measurements in 2D CFTs. In section \ref{section4} we adopt the OPE blocks as the quantum measurements for the QET protocol and calculate the energetics at each step. We first show that the QET will fail in the infinite time limit, and then show that the sub-leading correction beyond this limit will then yield QET by appropriate quantum feedback control. Finally, we give a toy example for demonstration of viable QET in CFT. We then conclude our paper in section \ref{section5} and end with a discussion on Bell inequality of the OPE blocks.

\section{Projection Measurements in CFT}\label{section2}

  A quantum measurement process can be described by a set of positive operators $\{ E_k \}$ whose sum is the identity operator, i.e.,
\be\label{POVM_1}
\sum_k E_k = \mathbb{I}\;.
\ee
Then, the probability of obtaining the outcome $k$ when measuring the state $|\psi\rangle$ is
\be\label{prob-k}
\textrm{p}_k=\langle \psi | E_k |\psi\rangle\;.
\ee
This is known as the positive operator-valued measure (POVM).  A special case is when the positive operators $E_k$'s are all projection operators, i.e., $E_k^{\dagger} E_j=\delta_{k,j} E_k$, then the normalized post-measurement state of outcome $k$ is
\be\label{post-m-s}
|\psi_k\rangle  = {E_k |\psi \rangle \over \sqrt{\langle \psi | E_k |\psi\rangle}}\;.
\ee
This is the so-called projective-valued measure (PVM).

   Moreover, the POVM can also be constructed by introducing the auxiliary probe coupled to the state $|\psi\rangle$, so that the operator  $E_k$ can be obtained as follows: acting on the total system by the time evolution operator  $U(t)$, and then projecting it onto the probe's eigenstate $|k\rangle_p$, i.e.,
\be
E_k:=M_k^{\dagger}M_k
\ee
with
\be
M_k:=\;_p\langle k| U(t) |0\rangle_p,
\ee
where the subscript $p$ denotes ``probe". It is easy to see that \eq{POVM_1} is satisfied by $U^{\dagger} U=1$.

    Based on the above procedure, one may construct the POVM in quantum field theory (QFT) and then implement them on some quantum tasks, see for example \cite{Hotta:2011xj} on constructing POVM of free QFT for quantum energy teleportation. However, in practical the construction of POVM for interacting QFT is not so straightforward due to nontrivial operator mixings.
\subsection{OPE block in CFT}

    Instead, in $d$-dimensional CFTs there is a set of projection operators constructed by the shadow operator formalism \cite{shadows}, and explicitly they are given by
\be\label{pvm-1}
P_k = {\Gamma(\Delta_k) \Gamma(d-\Delta_k) \over \pi^d \Gamma(\Delta_k-{d\over 2}) \Gamma({d\over 2}-\Delta_k)} \int D^d X\; \mathcal{O}_k(X) |0\rangle \langle 0| \tilde{\mathcal{O}}_k(X),
\ee
where $\Gamma(x)$ is the Gammas function and $\Delta_k$ is the conformal dimension of $\mathcal{O}_k$. These projectors are complete if $k$ runs over all primaries, i.e.,
\be
\sum_{k \in \textrm{all primaries}} P_k=  \mathbb{I}_{CFT}\;.
\ee
We have introduced the shadow operator 
\be\label{shadow op}
\tilde{\mathcal{O}}_k(X) := \int D^d Y \; {1\over (-2 X \cdot Y)^{d-\Delta_k} } \mathcal{O}_k(Y),
\ee
so that  it can be used to show that
\be\label{PVM-1}
P_iP_j= \delta_{i,j} P_i\;.
\ee
In the above, we adopt the notation of embedding space for the coordinate $X$, i.e., for CFT in d-dimensions, the ``embedding space" is ${\mathbb R}^{d,2}$. The dimensional space is obtained by quotienting the null cone $X^2=0$ and by the rescaling $X\sim \lambda X$, $\lambda \in {\mathbb R}$. In particular, we can choose the Poincare section such that $X:=(X^+,X^-,X^{\mu})=(1,x^{\mu}x_{\mu},x^{\mu})$ such that
\be
-2 X_1 \cdot X_2= (x_1-x_2)^2\;. \nn
\ee
The ``conformal integral" in \eq{shadow op} is defined by \cite{shadows}
\be
\int D^dX \; f(X) = {1\over \textrm{Vol} \; \textrm{GL}(1,\mathbb{R})^+} \int_{X^++X^-\ge 0}  d^{d+2}X \; \delta(X^2) \; f(X)\;. \nn
\ee

  Even though $P_j$'s are projection operators, however, it is not local and thus we cannot use them to implement local quantum measurements which are required in many quantum information tasks such as quantum (energy) teleportation. Fortunately, for 2D CFTs the $P_j$ becomes a quasi-local operator when acting on the following states
\be\label{quasilocal state}
O_1(x_1)O_2(x_2)|0\rangle,
\ee
where $|0\rangle$ is the ground state of CFT. In this case, the integration in \eq{pvm-1} and \eq{shadow op} is over the casual diamond ${\cal D}_{\bf A}$ subtended by the interval $[x_1,x_2]$, i.e., $x_1<x_2$ w.l.o.g..  For simplicity, we will only consider the case with $O_1=O_2:=O_i$ the primary operator of conformal dimension $(h_i, \bar{h}_i)$. We can then view the state \eq{quasilocal state} as some quasi-local excitation prepared by Alice, and then she further performs a local projection measurement within her causal domain for some quantum information task.

 Indeed, the post-measurement state is related to the OPE block defined in \cite{opeblock}, i.e.,
\be\label{ope-b-1}
P_k \; \mathcal{O}_i(x_1) \mathcal{O}_i(x_2) |0\rangle = x_{12}^{-2h_i-2\bar{h}_i} c_{i i k}\, \mathcal{B}_k(x_1, x_2) |0\rangle\;.
\ee
where $c_{i i k}$ is the OPE coefficient and $x_{mn}:=x_m-x_n$. By this definition, it is straightforward to relate the conformal block $g_k(u,v)$ and the two-point correlator of the OPE blocks, i.e.,
\bea
g_k(u,v)&:=& (x_{12} x_{34}) ^{2 h_i +2 \bar{h}_i} c^{-2}_{iik} \; \langle 0| \mathcal{O}_i(x_3) \mathcal{O}_i(x_4) \; P_k \; \mathcal{O}_i(x_1) \mathcal{O}_i(x_2) |0\rangle \label{CB2}\nonumber
\\
&=& \langle 0|\mathcal{B}^{\dagger}_k(x_4,x_3) \mathcal{B}_k(x_1,x_2)|0 \rangle,\label{CB1}
\eea
where the cross ratio $u$ is
\be
u:={x^2_{12} x^2_{34} \over x^2_{13} x^2_{24}}\;, v={x^2_{14} x^2_{23} \over x^2_{13} x^2_{24}}\;.
\ee
Note that the second equality is arrived by the definition \eq{ope-b-1} of the OPE block.

\bigskip

 In 2D Minkowski spacetime  the OPE block $\mathcal{B}_k(x_1, x_2)$ can be expressed in terms of a smearing operator over the causal diamond ${\cal D}_{\bf A}$, i.e.,
\be\label{OPE block}
\mathcal{B}_k(x_1,x_2)=\int_{\mathcal{D}_{\bf A}} d^2w  \; G_k(w,\bar{w};x_1,x_2) \mathcal{O}_k(w,\bar{w}),
\ee
where $O_k$ is a primary operator of conformal dimension $(h_k,\bar{h}_k)$. The smearing function $G_k(w,\bar{w};x_1,x_2)$ is the propagator constructed in the framework of the integral geometry \cite{opeblock}, and takes the following form in the large central charge limit:
\begin{eqnarray}\label{Gk-1}
G_k(w_0,\bar w_0; x_1,x_2)=n_k \bar{n}_k \Big(\frac{w_{01}w_{20}}{x_{21}}\Big)^{h_k-1}\Big( \frac{\bar{w}_{01}\bar{w}_{20}}{x_{21}}\Big)^{\bar h_k-1},
\end{eqnarray}
where the overall factors $n_k$ and $\bar{n}_k$ are
\be\label{nkk}
n_k:=\frac{\Gamma(2h_k)}{\Gamma( h_k)^2}\;, \qquad \bar{n}_k:=\frac{\Gamma(2\bar h_k)}{\Gamma(\bar h_k)^2} \;.
\ee
In the above and hereafter, we denote the lightcone coordinates of the spacetime point $(t,x)$ by
\be
w:=x-t\;, \qquad  \bar{w}:=x+t\;.
\ee
Especially, $w=\bar{w}=x$ on $t=0$ slice. We also introduce the short-handed notation: $w_{ij}:=w_i-w_j$ and $\bar{w}_{ij}:=\bar{w}_i-\bar{w}_j$.
\subsection{OPE block as POVM}
The form of (\ref{CB1}) is similar as the definition of the probability of POVM (\ref{prob-k}) if we take the limit $x_1\to x_4$ and $x_2 \to x_3$. This motives us to construct the POVM operators by OPE block with some suitable normalization and regularization.\\
From \eq{prob-k} the probability for the outcome $k$ is formally given by
\be
\textrm{p}_k = {c_{i i k}^2 g_k(1,0) \over  \sum_j  c_{i i j}^2 g_j(1,0) } \label{prob-pvm-1}\;, \qquad \mbox{with}\; \;\sum_k p_k=1
\ee
where $g_k(u,v)$ is the conformal block defined by \eq{CB1}. After the measurement, the outcome state becomes
\be\label{outcome-state}
|\psi_k\rangle:={\mathcal{B}_k(x_1,x_2)|0 \rangle \over \sqrt{g_k(1,0)}}\;.
\ee
Note that $p_k$ is independent of the value of $x_1$ and $x_2$ though $|\psi_k\rangle$ does.

  Using \eq{OPE block} and \eq{Gk-1} we can evaluate the universal global conformal blocks for 2D CFTs, and the results are
\begin{eqnarray}\label{gCBB}
g_k(u,v)=z^{h_k}\bar z^{\bar h_k} ~_2F_1(h_k,h_k,2h_k,z)~_2F_1({\bar h}_k,{\bar h}_k,2 {\bar h}_k,\bar z)\;,
\end{eqnarray}
where $z=w_{12}w_{34}/(w_{13}w_{24})$ and $\bar z=\bar w_{12}\bar w_{34}/(\bar w_{13}\bar w_{24})$ are the cross ratios, and $u=z\bar z, v=(1-z)(1-\bar z)$.


In the limit $x_1\to x_4, x_2\to x_3$ , $z,\bar z\to 1$. Notice that for $h_k, {\bar h}_k >0$
\be
g_k(u\to 1,v\to 0)\to  ~_2F_1(h_k,h_k,2h_k,1)~_2F_1({\bar h}_k,{\bar h}_k,2 {\bar h}_k,1)\;,
\ee
which is formally divergent and needs some regularization. Moreover, it is easy to see that $g_k(1,0)$ should be dimensionless, and thus the divergence is log divergence. In fact, by definition of \eq{outcome-state}, this regularization can be understood as the wavefuntion renormalization. At this moment we only formally treat $g_{k \ne 0}(1,0)$ as a regularized function of running energy scale $\mu$ in the form of $\log{\Lambda \over \mu}$ where $\Lambda$ is some UV cutoff energy scale. We will discuss more details on regularization in next subsection. Obviously, the particular smearing function \eq{Gk-1} helps to avoid the more serious divergence such as the power-law ones.

On the other hand, for the vacuum/identity global conformal block denoted by $k=0$ with $h_0={\bar h}_0=0$ we can check
\be
g_0(1,0)=1\;.
\ee
For $k,k'\ne 0$, the ratio $g_k(u,v)/g_{k'}(u,v)$ is finite in the limit $u\to 1,v\to 0$, i.e.,
\begin{eqnarray}
\lim_{u\to 1, v\to 0} \frac{g_k(u,v)}{g_{k'}(u,v)}=\frac{n_k\bar n_k}{n_{k'}\bar n_{k'}},
\end{eqnarray}
which is only related to the conformal dimensions according to the definition (\ref{nkk}) of $n_k$ and $\bar n_{k}$.\\
From the above we can conclude that
\be\label{probability}
\textrm{p}_0\simeq 0, \qquad \textrm{p}_{k\ne 0}= {c^2_{i i k} n_k \bar{n}_k \over \sum_{j \ne 0} c^2_{i i j} n_j \bar{n}_j}
\ee
where $\simeq$ means that the equality holds exactly if the UV cutoff is taken to infinity. Since $\Gamma(a>0)>0$, thus $\textrm{p}_k$ should be positive and finite for all $h_k, {\bar h}_k >0$, as expected.  It is interesting to see $\textrm{p}_0\simeq 0$ so that the identity channel is excluded as a physical outcome.

 Finally, based on all the above, we can write down the density matrix $\rho_{\bf A}$ for the resultant state  after Alice's quasi-local measurement on the state \eq{quasilocal state}, i.e.,
\be
\rho_{\bf A} =\sum_k \textrm{p}_k |\psi_k\rangle \langle \psi_k|=\sum_{k \ne 0}  {c_{i i k}^2  \mathcal{B}_k(x_1,x_2) |0\rangle\langle 0| \mathcal{B}^{\dagger}_k(x_1,x_2) \over  \sum_{j \ne 0}  c_{i i j}^2 g_j(1,0) } \;.
\ee

We can also introduce the POVM-like measurement operator
\be\label{mk}
M_k({\bf A}):=\sqrt{\textrm{p}_k \over g_k(1,0)} \; \mathcal{B}_k(x_1,x_2)
\ee
so that $\rho_{\bf A}$ can be expressed as
\be \label{post-m-state}
\rho_{\bf A} := \sum_{k\ne 0} M_k({\bf A}) |0\rangle \langle 0| M^{\dagger}_k (\bf A)\;.
\ee
Thus,
\be\label{pk-Mk}
\textrm{p}_k= \langle 0| M^{\dagger}_k {\bf A}) M_k({\bf A}) |0\rangle\;.
\ee
The above form of $\rho_{\bf A}$ suggests that we can also think that all the non-vacuum OPE blocks, i.e., $M_{k\ne 0}({\bf A})$, form a complete set of non-trivial measurement operators when acting on the CFT ground state. After the measurement, the CFT ground state then get quasi-locally excited so that the vacuum OPE block is excluded, i.e., $\textrm{p}_0=0$ but $\textrm{p}_k\ne 0$.  However, the set $\{ M_{k\ne 0}({\bf A}) \}$ cannot be complete if it does not act on the CFT ground state just because  in the operator sense
\be
\sum_{k\ne 0} M^{\dagger}_k({\bf A}) M_k({\bf A})=\sum_{k\ne 0}\; {\textrm{p}_k \over g_k(1,0)} \; O_i(x_2) O_i(x_1) P_k O_i(x_1) O_i(x_2) \ne \mathbb{I}\;,
\ee
i.e., the completeness does not hold for arbitrary states.

\subsection{On regularization}

As mentioned in last subsection we meet with a divergent quantity $g_{k\ne 0}(1,0)$. Indeed when dealing with quantum field theory one often meets with the UV-divergence. But we expect   we could obtain some physical quantities which are independent of UV cut-off  by suitable regularization, such as the $S-$matrix in a scattering process. Here our definition of probability is similar.\\

Our starting point is the state (\ref{quasilocal state}), which is a local state in the sense that the energy density of this state is divergent at point $x_1$ and $x_2$. This state is not normalizable. One could regulate it by moving the operator slightly into Euclidean time, i.e, at $t=i\delta$, where $\delta$ is a small positive number. We could define a new state

\begin{eqnarray}
e^{-\delta H}O_1(x_1)O_2(x_2)|0\rangle,
\end{eqnarray}
where $H$ is the Hamiltonian of CFT. This method is first used in \cite{Calabrese:2006rx} to discuss quantum quench by a boundary state. Therefore we could define the post-measurement state (\ref{outcome-state}) as
\begin{eqnarray}\label{new2}
\ket{\psi_k}={N_k(\delta)}^{-1} e^{-\delta H}{\mathcal{B}_k(x_1,x_2)|0 \rangle }\;,
\end{eqnarray}
the normalization constant is related to the parameter $\delta$, which is regarded as a UV cut-off. After some lengthy calculations, we could obtain
\begin{eqnarray}
N^2_k(\delta)= ~_2F_1(h_k,h_k,2h_k,1-\frac{\delta^2}{L^2})~_2F_1({\bar h}_k,{\bar h}_k,2 {\bar h}_k,1-\frac{\delta^2}{L^2})\;,
\end{eqnarray}
where $L:=|x_{12}|$. Note that $N_k\sim \log{\delta/L}$ if $\delta \ll L$. This process actually regularize $g_k(1,0)$ by $N^2_k$. Therefore we get the regularized POVM-like operator
\be\label{mk1}
M_k({\bf A}):=\frac{\sqrt{\textrm{p}_k}}{N_k(\delta)} \; e^{-\delta H}\mathcal{B}_k(x_1,x_2).
\ee
The regularization make the post-measurement state (\ref{new2}) be a normalizable state. But the probability is still not dependent upon the UV cut-off as long as $\delta/L\ll 1$.

\section{QET protocol in 2D CFTs}\label{section4}

   Based on the above construction of the measurement process for 2D CFTs, we are now ready to consider a corresponding QET protocol. The protocol goes as follows. First, Alice performs the projection measurement $\{ P_k \}$ of \eq{pvm-1} on the quasi-local excited state \eq{quasilocal state}, and send her measurement outcome to distant Bob via classical communication (CC). According to the outcome, Bob then perform the following quasi-local unitary operation (LO) on the interval $[x_3,x_4]$ which is far from the interval $[x_1,x_2]$:
\begin{eqnarray}
U_{\bf B}=e^{i \beta_k \theta G_{\bf B}}\\ \nonumber
\end{eqnarray}
where  $\beta_k$ is the feedback-control parameter associated with outcome $k$, $\theta$ labels the angle for unitary transformation, and we choose
\begin{eqnarray}
G_{\bf B}=\int_{x_4}^{x_3}dx \; f(x) \; O_h (t_0,x),
\end{eqnarray}
with  $t_0$ some constant time, $O_h$ being some primary operator of conformal weight $(h,\bar h)$, and $f(x)$  a smooth smearing real function. Thus $G_{\bf B}$ is hermitian so that $U_{\bf B}$ is unitary.   In QET, one can tune $\beta_k$ to help Bob extract energy by local operations.

 Based on the above QET protocol with a many-body entangled state as the resource of the task, and with the help of LOCC, we will perform the energetic analysis for each step in the following.

 We start with Alice's post-measurement state, which is already given by \eq{post-m-state}.  Thus, the energy injected by the projection measurement is
\be\label{EA-1}
E_{\bf A}=\tr(\rho_{\bf A} H_{CFT})=\int dx \sum_{k\ne 0} \langle 0| M^{\dagger}_k({\bf A}) T_{00}(x) M_k(\bf A)|0\rangle\;.
\ee
where $H_{CFT}:=\int dx \; T_{00}(x)$ is the Hamiltonian of CFT. We can manipulate \eq{EA-1} further to get some explicit form of $E_{\bf A}$. However, the calculation is difficult to carry out to the end due to the complication of triple integration $\int dx \int d^2w \int d^2w' \cdots$, and the final form is irrelevant to the viability of QET as shown below. Thus, we will not pursue this further.

   Assume the time elapses $T$ before Bob performs the local operation, then in the Heisenberg's picture the measurement operator $M_k(\bf A)$ evolves into $M_k({\bf C}):=e^{-i H_{CFT} T} M_{k}({\bf A})e^{i H_{CFT} T}$, or more explicitly,
\be\label{MkC}
M_k({\bf C})=\frac{\sqrt{p_k}}{ N_k(\delta)} \int_{\cal{D}_{\bf A}} d^2w\; G_k(w,\bar w;x_1,x_2) e^{-\delta H}O_k(w-T,\bar w+T)\;.
\ee
After Bob's local unitary operation the total state of CFT becomes
\be
\rho_{QET}=\sum_{k\ne 0} U_{\bf B} M_k({\bf C}) |0 \rangle \langle 0| M^{\dagger}_k({\bf C}) U^{\dagger}_{\bf B}\;.
\ee
We can then evaluate the amount of energy teleported from Alice to Bob as following:
\be
E_{\bf B}=E_{\bf A}  -\sum_{\k\ne 0}  \langle 0| M^{\dagger}_k({\bf C}) U^{\dagger}_{\bf B} H_{CFT}  U_{\bf B} M_k({\bf C}) |0 \rangle.
\ee

   If we assume $\theta$ is small, then we can express $E_{\bf B}$ in in terms of $\theta$ expansion, i.e.,
 \begin{eqnarray}\label{EB-theta}
 &&E_{\bf B}= i\theta \sum_{k\ne 0} \beta_k  \langle 0| M^{\dagger}_k({\bf C})[H_{CFT}, G_{\bf B}] M_k({\bf C}) |0 \rangle \nonumber \\
 &&\quad \quad -  {\theta^2\over 2} \sum_{k\ne 0} \beta_k^2  \langle 0| M^{\dagger}_k({\bf C})  [[H_{CFT}, G_{\bf B}], G_{\bf B}] M_k({\bf C}) |0 \rangle  +\cdots
\end{eqnarray}
where we have used the fact that
\be\nn
\sum_{k\ne0}\langle 0| M^{\dagger}_k({\bf C}) H_{CFT}  M_k({\bf C}) |0 \rangle=\sum_{k\ne 0} \langle 0| M^{\dagger}_k({\bf A}) e^{i H_{CFT} T} H_{CFT}  e^{-i H_{CFT} T} M_k({\bf A}) |0 \rangle=E_{\bf A}\;.
\ee

 The commutator with $H_{CFT}$ in the above can be reduced to time-derivative by Heisenberg equation, i.e.,
\be\label{Hevo}
[H_{CFT}, O_h(\xi,{\bar \xi})] =-i(\partial_{\xi}-\partial_{\bar \xi} ) O_h(\xi,{\bar \xi})
\ee

\subsection{No-go in the infinite time limit} \label{section41}

  In this subsection, we will show the impossibility of having QET energy gain in the infinite time limit, i.e., $T\to \infty$.
To calculate the first order term of \eq{EB-theta}, denoted by $E_{\bf B}|_{\theta}$,  for each $k\ne 0$ term we need to deal with
\bea
E^{(1)}_k&:=& i\beta_k\theta \langle 0| M^{\dagger}_k({\bf C})[H_{CFT}, G_{\bf B}] M_k({\bf C}) |0 \rangle \nn \\
&=& i\beta_k\theta \int_{x_3}^{x_4} dx \;  f(x) \; \langle 0| M^{\dagger}_k({\bf C})[H_{CFT}, O_h (\xi,{\bar \xi})] M_k({\bf C}) |0 \rangle
\eea
where we have introduced the lightcone coordinates $\xi:=x-t_0$ and ${\bar \xi}:=x+t_0$.\\
 By using (\ref{MkC}) and (\ref{Hevo}) we have
\begin{eqnarray}
&&\langle 0| M^{\dagger}_k({\bf C})[H_{CFT}, O_h (\xi,{\bar \xi})] M_k({\bf C}) |0 \rangle \nn \\
&&=-i(\partial_{\xi}-\partial_{\bar \xi} ) \langle 0| M^{\dagger}_k({\bf C}) O_h (\xi,{\bar \xi}) M_k({\bf C}) |0 \rangle \\
&&=\frac{-i p_k}{ N^2_k(\delta)} \int_{\cal{D}_{\bf A}} \int_{\cal{D}_{\bf A}}d^2wd^2w'\; G_k(w,\bar w;x_1,x_2)\; G_k(w',\bar w';x_1,x_2)\\ \nonumber
 && \times (\partial_{\xi}-\partial_{\bar \xi} )\langle  O_k(w-T-i\delta,\bar w+T+i\delta)O_h (\xi,{\bar \xi}) O_k(w'-T+i\delta,\bar w'+T-i\delta)\rangle,
\end{eqnarray}
which is related to the three point correlation function $\langle O_k(w-T,\bar w+T)O_h (\xi,{\bar \xi}) O_k(w'-T,\bar w'+T)\rangle$,
\begin{eqnarray}\label{new1}
&&\langle O_k(w-T-i\delta,\bar w+T+i\delta)O_h (\xi,{\bar \xi}) O_k(w'-T+i\delta,\bar w'+T-i\delta)\rangle\nonumber \\
&&\propto\frac{1}{(\xi-w+T-i\delta)^{h}(\xi-w'+T+i\delta)^{h}(w-w'-2i\delta)^{2h_k-h}} \nn \\
&&\quad \times\frac{1}{(\bar \xi-\bar w-T+i\delta)^{\bar h}(\bar \xi-\bar w'-T-i\delta)^{\bar h}(\bar w-\bar w'+2i\delta)^{2\bar h_k-\bar h}},
\end{eqnarray}
which vanishes in the infinite time limit if $h\ne 0$. In short, this implies that
\be\label{clustering}
\lim_{T\to \infty} \; \langle 0| M^{\dagger}_k({\bf C}) O_{h\ne 0} (\xi,{\bar \xi}) M_k({\bf C}) |0 \rangle=0\;.
\ee
Thus, $E^{(1)}_{k\ne 0}$ vanishes in the infinite time limit if $h\ne 0$. This yields the fact that there is no QET energy gain or loss at the first order of $\theta$ expansion if taking the limit $T\to \infty$.

 Since $E_{\bf B}|_{\theta}=0$ in the infinite time  limit, we now go to evaluate the second order term of $E_{\bf B}$, denoted as $E_{\bf B}|_{\theta^2}$, i.e.,
\be \label{Etheta2}
E_{\bf B}|_{\theta^2}:={\theta^2 \over 2} \sum_{k\ne0} E^{(2)}_k=-  {\theta^2\over 2} \sum_{k\ne 0} \beta_k^2 \; \langle 0| M^{\dagger}_k({\bf C})  [[H_{CFT}, G_{\bf B}], G_{\bf B}] M_k({\bf C}) |0 \rangle\;.
\ee
Similar to simplification for the $E^{(1)}_{k}$, by using \eq{Hevo} we can express $E^{(2)}_{k}$ as following:
\bea
E^{(2)}_{k} =i \beta^2_k  \int_{x_3}^{x_4} dy_1 f(y_1)\int_{x_3}^{x_4} dy_2  f(y_2)\;  (\partial_{\xi_1} -\partial_{\bar{\xi}_1}) \langle 0| M^{\dagger}_k({\bf C}) [O_h(\xi_1,\bar{\xi}_1),O_h(\xi_2,\bar{\xi}_2)] M_k({\bf C}) |0\rangle \qquad\nn
\eea
where we have introduced $\xi_i:=y_i-t_0$ and $\bar{\xi}_i:=y_i+t_0$ for $i=1,2$. Using the definition of $M_k({\bf C})$, the correlator inside the above double integral can be further expressed in terms of four-point function  $\langle O(w-T,\bar w+T)O_h(x) O_h(y) O(w'-T,\bar w'+T)\rangle$. We can further reduce this into the sum of  three-point functions by the OPE of $O_h(x) O_h(y)$. Using the fact of \eq{clustering} in the long time limit and also \eq{pk-Mk}, we can arrive
\be
\lim_{T\to \infty} \; \langle 0| M^{\dagger}_k({\bf C}) O_{h} (\xi_1,{\bar \xi}_1) O_h(\xi_2,{\bar \xi}_2) M_k({\bf C}) |0 \rangle= {c_{hh0} \; p_k \over (\xi_1-\xi_2)^{2h} (\bar{\xi}_1-\bar{\xi}_2)^{2\bar{h}} }
\ee
where  $c_{hh0}$ is the OPE coefficient for the identity channel.  If we recognize the above power-law factor obtained from OPE as the two-point function $ \langle 0|O_{h} (\xi_1,{\bar \xi}_1) O_h(\xi_2,{\bar \xi}_2)|0\rangle$, then we find that the 4-point function is cluster-decomposed in the infinite time limit.  This then implies that
\begin{eqnarray}
&&\lim_{T\to \infty} \; (\partial_{\xi_1} -\partial_{\bar{\xi}_1}) \langle 0| M^{\dagger}_k({\bf C}) [O_{h} (\xi_1,{\bar \xi}_1), O_h(\xi_2,{\bar \xi}_2)] M_k({\bf C}) |0 \rangle \\ \nonumber
&&\quad \quad \quad = c_{hh0}\; p_k\; (\partial_{\xi_1} -\partial_{\bar{\xi}_1}) \langle 0|  [O_{h} (\xi_1,{\bar \xi}_1), O_h(\xi_2,{\bar \xi}_2)]|0 \rangle \; .
\end{eqnarray}

As $O_h(\xi_1,\bar \xi_1)$ and $O_h(\xi_2,\bar \xi_2)$ are operators on time slice $t=t_0$, naively the commutator seems to be zero except $\xi_1=\xi_2$. However, due to the overall derivative on the commutator, one should shift the coordinate away for the slice $t=t_0$ to make it well-defined. So generally this would be a non-zero result.

To see the sign of $E_k^{(2)}$ let's turn to Fourier space. To carry out the calculations, we assume the x-direction to be periodic with $x\sim x+L$. By  a coordinate transformation in the Euclidean space
\be
z^E=e^{-2\pi i \xi^E/L},
\ee
the cylinder is mapped to infinite plane, on which the Euclidean correlator is
\be
\langle O(z^E_1,\bar z^E_1)O(z^E_1,\bar z^E_2)\rangle=\frac{1}{(z^E_1-z^E_2)^{2h}(\bar z^E_1-\bar z^E_2)^{2\bar h}}.
\ee
The correlator on the cylinder can then be obtained as
\be
\langle O(\xi^E_1,\bar \xi^E_1)O(\xi^E_2,\bar \xi^E_1)\rangle =(\frac{2\pi}{L})^{2h}(\frac{2\pi}{L})^{2\bar h}\frac{e^{2\pi i h(\xi^E_1-\xi^E_2)/L}e^{-2\pi i h(\bar \xi^E_1-\bar\xi^E_2)/L}}{(1-e^{2\pi i (\xi^E_1-\xi^E_2)/L})^{2h}(1-e^{-2\pi i (\bar \xi^E_1-\bar \xi^E_2)/L})^{2\bar h}}.
\ee
By the $i\epsilon$ prescription\cite{Hartman:2015lfa}, the corresponding Minkowski correlators are
\be
\langle O(\xi_1,\bar \xi_1)O(\xi_2,\bar \xi_2)\rangle=(\frac{2\pi}{L})^{2h}(\frac{2\pi}{L})^{2\bar h}\frac{e^{2\pi i h(\xi_1-\xi_2)/L}e^{-2\pi i h(\bar\xi_1-\bar \xi_2)/L}}{(1-e^{2\pi i (\xi_1-\xi_2)/L-\epsilon})^{2h}(1-e^{-2\pi i (\bar \xi_1-\bar \xi_2)/L-\epsilon})^{2\bar h}},
\ee
and
\be
\langle O(\xi_2,\bar \xi_2)O(\xi_1,\bar \xi_1)\rangle =(\frac{2\pi}{L})^{2h}(\frac{2\pi}{L})^{2\bar h}\frac{e^{2\pi i h(\xi_1-\xi_2)/L}e^{-2\pi i h(\bar\xi_1-\bar \xi_2)/L}}{(1-e^{2\pi i (\xi_1-\xi_2)/L+\epsilon})^{2h}(1-e^{-2\pi i (\bar \xi_1-\bar \xi_2)/L+\epsilon})^{2\bar h}},
\ee
where $\epsilon$ is a small positive number. Thus we could expand the above expressions as
\be
\langle O(\xi_1,\bar \xi_1)O(\xi_2,\bar \xi_2)\rangle =(\frac{2\pi}{L})^{2h}(\frac{2\pi}{L})^{2\bar h}\sum_{n\ge 0}\sum_{m\ge 0}F_n(h)F_{m}(h)e^{2\pi i(\xi_1-\xi_2)(n+h)/L}e^{-2\pi i(\bar \xi_1-\bar \xi_2)(m+\bar h)/L},
\ee
and
\be
\langle O(\xi_2,\bar \xi_2)O(\xi_1,\bar \xi_1) \rangle=(\frac{2\pi}{L})^{2h}(\frac{2\pi}{L})^{2\bar h}\sum_{n\ge 0}\sum_{m\ge 0}F_n(h)F_{m}(h)e^{-2\pi i(\xi_1-\xi_2)(n+h)/L}e^{2\pi i(\bar \xi_1-\bar \xi_2)(m+\bar h)/L},
\ee
where $F_{n}$ are the coefficients of the Tayor series of $(1-x)^{-2h}$, which are positive definite. Now we make the coordinates $(\xi_1,\bar \xi_1)$ and $(\xi_2,\bar \xi_2)$ approach to the time slice $t=t_0$, then   $E_k^{(2)}$ becomes
\be\label{secondordercluster}
E_k^{(2)}=-\beta_k^2\; c_{hh0}\; p_k \; (\frac{4\pi}{L})(\frac{2\pi}{L})^{2h}(\frac{2\pi}{L})^{2\bar h}
\sum_{n\ge 0}\sum_{m\ge 0}F_n(h)F_{m}(h)(m+n+h+\bar h) |f_{n-m}(h)|^2\;,
\ee
where we have defined
\be
f_{n-m}=\int_{x_3}^{x_4}dy \; f(y) \; e^{2\pi i(n-m+h-\bar h)}\;. \nn
\ee
From (\ref{secondordercluster}) we could see $E_k^{(2)}$ is negative definite, and thus $E_{\bf B}|_{\theta^2}$ is always negative.  This means that Bob cannot extract energy  in the infinite time limit up to second order.

\subsection{Sub-leading correction beyond long time limit}\label{section42}

  We now try to consider the sub-leading correction of $E_{\bf B}|_{\theta}$ in the large $T$ expansion.   We will see that there are nonzero energy changes for each channel so that we can manipulate the feedback control parameters to obtain QET. Thus, the no-go theorem is lifted beyond the long time limit.

  Without taking the long time limit, we should deal with $E^{(1)}_k$ in the following form:
\be\label{Ek1}
E^{(1)}_k= \beta_k\theta \int_{x_3}^{x_4} dx \; f(x) ( \partial_{\xi}-\partial_{\bar \xi} ) \langle 0| M^{\dagger}_k({\bf C})O_h (\xi,{\bar \xi}) M_k({\bf C}) |0 \rangle \;.
\ee
For simplicity, hereafter we will set $f(x)=1$.

We can either calculate the 3-point function in \eq{Ek1} directly, or we can just consider the sub-leading contribution in the large $T$ expansion. Both yields the similar results, and for simplicity we will just consider the latter \footnote{As we can see from (\ref{new1}) the integration  should also depend on the cut-off $\delta$, here we only consider the leading contributions, so ignore $\delta$ in the integration. },
\be\label{E1k-int}
E^{(1)}_k \simeq  {\cal N}_{\theta,k,T} \int_{\cal D_{\bf A}} d^2w_0 \int_{\cal D_{\bf A}} dw^2_3 \; G_k(w_0,\bar{w}_0;x_2,x_1) G_k(w_3,\bar{w}_3;x_1,x_2) {1\over w_{03}^{2h_k-h} \bar{w}_{03}^{2\bar{h}-\bar{h}_k}}
\ee
with the overall factor
\be
{\cal N}_{\theta,k,T}:=- \beta_k \; \theta\; { \textrm{p}_k \over N^2_k(\delta) } {2\Delta_h \over T^{2\Delta_h+1}}  \; w_{21}
\ee
where $\Delta_h:=h+\bar h$. \\
 Plugging (\ref{Gk-1}) into (\ref{E1k-int})
\be\label{E1k-int-simplify}
E^{(1)}_k \simeq  {\cal N}_{\theta,k,T} n_k \bar n_k I_{03}\bar I_{03},
\ee
where we define
\be\label{integral}
I_{03}=\int_{w_1}^{w_2}dw_0 \int_{w_1}^{w_2}dw_3 \; (w_{01}w_{20})^{h_k-1} (w_{31}w_{23})^{h_k-1} w_{03}^{h-2h_k},
\ee
and $\bar I_{03}$ is the anti-holomorphic part. If $h-2h_k$ is not an integer, the integral will have branch cut. To simplify the calculation we assume $h-2h_k$ is an integer, and constrain $h-2h_k>-1$. In this case we could obtain an analytic result of the integral $I_{03}$. After some calculations, it is given by
\begin{eqnarray}\label{integral03}
&&I_{03}=w_{21}^{h+2h_k-2}(1+(-1)^{h-2h_k})\frac{\sqrt{\pi}}{2^h}\frac{\Gamma(h_k)^2\Gamma(h-2h_k+1)\Gamma(\frac{h}{2})}{\Gamma(\frac{h+1}{2})\Gamma(1+\frac{h}{2}-h_k)
\Gamma(\frac{h}{2}+h_k)},
\end{eqnarray}
which is positive or zero in the region $h-2h_k>-1$.  $\bar I_{03}$ can be obtained by  $h\to\bar h$ and $h_k\to\bar h_k$ in expression (\ref{integral03}). Plugging them into (\ref{E1k-int-simplify}) we get the final result of $E^{(1)}_k$ , which could be positive by tuning parameter $\beta_k$. The first non-vanishing contribution to Bob's extraction energy beyond the infinite time limit is then
\be\label{firstorder}
E^{(1)}_{\bf{B}}:=\sum_{k\ne 0}E^{(1)}_{k},
\ee
where $E^{(1)}_k$ is given by (\ref{E1k-int-simplify}).

   Finally, we remark about requiring $h-2h_k$ to be some integer in the above discussion. This is assumed to avoid the branch cut for the integral related to the integral representation of hypergeometric function. Supposed that we do not restrict to the integer values of $h-2h_k$, then we need to perform suitable analytic continuation to carry out the integration, and it may result in a complex-valued energy, i.e., complex $E_k^{(1)}$. Physically, the complex energy is expected as the quasi-local states such as $M_k({\bf C}) |0 \rangle$ or $U_{\bf B} M_k({\bf C}) |0 \rangle$ are not energy eigenstates, i.e., the states may not be stable under evolution. This then implies that these states with non-integer $h-2h_k$ may be quasi-normal states with the imaginary part of the energy as their decay width.

\subsection{QET in a toy 2D CFT model}\label{subsectioneg}
In this subsection we will use a toy example to show our previous abstract discussion of viable QET. Assume in this model the OPE of $\mathcal{O}_i(x_1)\mathcal{O}_i(x_2)$ only has two channels except the identity channel, i.e.,
\be
\mathcal{O}_i(x_1)\mathcal{O}_i(x_2)=x_{12}^{-2h_i-2\bar{h}_i} \sum_{k \in \{0,1,2\}}  c_{i i k}\, \mathcal{B}_k(x_1, x_2)\;,
\ee
where $k=0$ refers to the identity, $1,2$ are two others. Without loss of generality, we normalize $c_{ii1}=1$, denote $c_{ii2}=c$. According to (\ref{probability}) we have
\be
p_0=0, \quad \quad p_1=\frac{n_1\bar n_1}{n_1\bar n_1+c n_2\bar n_2},\quad \quad p_2=\frac{c n_2\bar n_2}{n_1\bar n_1+c n_2\bar n_2},
\ee
where $n_i=\frac{\Gamma(2h_i)}{\Gamma(h_i)^2}$ and  $\bar n_i=\frac{\Gamma(2\bar h_i)}{\Gamma(\bar h_i)^2}$$(i=1,2)$. By using (\ref{firstorder}), we obtain the energy Bob can extract $E^{(1)}_{\bf{B}}$,
\be
E^{(1)}_{\bf{B}}= -C \frac{1}{T^{2(h+\bar h)+1}}\sum_{k=1,2}\beta_k p_k M_k \bar M_k ,
\ee
where $C$ is positive constant unrelated to $k$, and
\begin{eqnarray}
&&M_k:=(1+(-1)^{h-2h_k})\frac{\Gamma(2h_k)\Gamma(h-2h_k+1)}{\Gamma(1+\frac{h}{2}-h_k)\Gamma(\frac{h}{2}+h_k)},\nonumber \\
&&\bar M_k:=(1+(-1)^{ \bar h-2\bar h_k})\frac{\Gamma(2\bar h_k)\Gamma(\bar h-2\bar h_k+1)}{\Gamma(1+\frac{\bar h}{2}-\bar h_k)\Gamma(\frac{\bar h}{2}+\bar h_k)}.
\end{eqnarray}
Let's do some numerical calculation. Assume $h_1=\bar h_1=1$ and $h_2=\bar h_2=\frac{3}{2}$, we need  $h>2$ since the constraint $h-2h_k>-1$. When $h=2 n$ ($n\ge 2$ and $n\in \mathcal{Z}$), we have
\be
E^{(1)}_{\bf{B}}=-C \frac{1}{T^{4n+1}}\beta_1 \frac{1}{1+c \frac{64}{\pi^2}} \Big(\frac{\Gamma(2n-1)}{\Gamma(n)\Gamma(n+1)}\Big)^2.
\ee
Therefore, in this case as long as  taking $\beta_1<0$, we will have $E^{(1)}_{\bf{B}}>0$, which means that Bob can extract energy by the unitary operation. We also notice that the result is proportional to $1/T^{2(h+\bar h +1)}$, so we should not use too large $h$ to ensure the energy $E^{(1)}_{\bf{B}}$ will not decay too fast. Also note that the decay behavior of $T$ is independent of $h_k$.

\section{Conclusion and Discussion}\label{section5}

 In comparison with the system of finite degrees of freedom, defining the quantum measurement process in QFT is a challenging problem, especially for the ones with non-trivial interactions. An obvious difficulty is the UV-divergence when the measurement operators contact with each others at the same spacetime point. But we usually expect that some suitable regularization methods could deal with this and help us to define physical quantities, which should be independent with UV-cutoff.

In this paper we make a modest step towards this problem. Our starting point is the shadow operator $P_k$ given in  (\ref{pvm-1}), which was once used to study the conformal blocks in Euclidean CFT. These operators $P_k$, which are complete, can be taken as projection measurements in CFT. When these operators work on a local state $\mathcal{O}_1(x_1) \mathcal{O}_2 (x_2)\ket{0}$, we could obtains a set of quasi-local states in 2D CFT. These state are obtained by the respective smearing operators, the so-called OPE blocks  $\mathcal{B}_k(x_1,x_2)$  over the casual diamond $\mathcal{D}$ of the interval $[x_1,x_2]$.

  We then use this operators to proceed the QET protocol in 2D CFT. We prove a no-go theorem if taking the long time limit. We calculate the energy that Bob could gain  up to second order of $\theta$, and find it impossible for Bob to gain any energy. This result is physical. As we know the entanglement of the state shared by Alice and Bob is the key for QET protocol's success.
Our result implies that the infinite time evolution will somehow destroy the entanglement resources shared between Alice and Bob for successful QET. This is also consistent with or due to the observed cluster decomposition in the infinite time limit. By considering the correction in the finite time duration we successfully realize the QET protocol. The energy Bob can extract depends on the UV-cutoff, which is different from the probabilities. This is reasonable  because the input energy by Alice is expected to be dependent with UV-cutoff.

   Before we close our paper in this section, we will comment on the issue of checking Bell inequality by using our weak-sense POVM-like operators. Bell inequality formulated in the CHSH form needs two pairs of Hermitian operators, say $A_1,A_2$ for Alice and $B_1,B_2$  for Bob, the norms of which are required to be smaller than one. We assume $A_1:=M_{k_1}(\bf{A})$, $A_2:=M_{k_2}(\bf{A})$, $B_1:=M_{k_2}(\bf{B})$, and $B_2:=M_{k_1}(\bf{B})$ with $k_1\ne k_2$. Note that the norms of these operators  in the vacuum state are all smaller than one as we can see from (\ref{pk-Mk}).
CHSH inequality is
\be
\gamma:=|\langle A_1(B_1+B_2)+ A_2(B_1-B_2)\rangle|\le 2.
\ee
If existing some operators such that the inequality is violated, we could claim the state has quantum entanglement. Let us see whether our measurement operators could make this. According to the definition of $M_k$ we have
\be
\langle 0|M_{k_1}({\bf{A}}) M_{k_2}(\bf{B})|0\rangle=0.
\ee
Therefore we arrive
\be
\gamma= \langle 0|M_{k_1}({\bf{A}}) M_{k_1}({\bf{B}})|0\rangle+\langle 0| M_{k_2}({\bf{A}})  M_{k_2}(\bf{B})|0\rangle.
\ee
If $x_3- x_2=L\ne 0$, i.e., the interval $[x_1,x_2]$ of Alice and the $[x_3,x_4]$ of Bob are separate, then $\langle 0|M_{k_1}({\bf{A}}) M_{k_1}({\bf{B}})|0\rangle$ will vanish if taking the UV-cutoff to zero. This can be seen as follows. Since $M_{k}= \frac{\sqrt{p_k}}{N_k(\delta)}\mathcal{B}_k$ so that
\be
\langle 0|M_{k_1}({\bf{A}}) M_{k_1}({\bf{B}})|0\rangle= \frac{p_{k_1}}{N^2_k(\delta)}\langle 0| \mathcal{B}_{k_1}({\bf{A}})\mathcal{B}_{k_1}({\bf{B}})|0\rangle=p_{k_1} \frac{g_{k_1}(u,v)}{N^2_k(\delta)},
\ee
where $u<1$ is the cross ratio. As $g_{k_1}(u,v)$ is finite, thus $\gamma$ will approach to zero. On the other hand,  if $L=0$ we will have $u\to 1$, then $g_{k_1}(u,v)\to N^2_k(\delta)$, $\gamma \to p_{k_1}+p_{k_2}\le 2$. In conclusion, the measurement operators $M_k$'s adopted here cannot violate the Bell inequality. \\

    Overall, our study implies that QET is viable in CFTs and can be used to detect the entanglement of the underlying quantum state even the corresponding Bell inequality using the same set of weak-sense measurement operators is not violated. Besides, we also point out many subtle issues regarding the quantum measurements in CFTs, which should deserve further investigations.

\subsection*{Acknowledgements}
We thank B. Czech for inspiring discussions on the initiation of this work.  We are appreciated for J. W. Chen, C. S. Chu, D. Giataganas, M. Hotta and J. j. Zhang for helpful discussions. FLL is supported by Taiwan Ministry of Science and Technology through Grant No.~103-2112-M-003-001-MY3, No.~106-2112-M-003-004-MY3 and No.~103-2811-M-003-024. He would also like to thank TAPIR and IQIM of Caltech for host his on leave from NTNU, during which part of this work is finished.

\providecommand{\href}[2]{#2}\begingroup\raggedright\endgroup


\begin{thebibliography}{1}

\bibitem{Ryu:2006bv}
  S.~Ryu and T.~Takayanagi,
  ``Holographic derivation of entanglement entropy from AdS/CFT,''
  Phys.\ Rev.\ Lett.\  {\bf 96}, 181602 (2006)
  [hep-th/0603001].

\bibitem{Ryu:2006ef}
  S.~Ryu and T.~Takayanagi,
  ``Aspects of Holographic Entanglement Entropy,''
  JHEP {\bf 0608}, 045 (2006)
  [hep-th/0605073].

\bibitem{spt} X. Chen, Z.-C. Gu, Z.-X. Liu and X.-G. Wen, ``Symmetry protected topological orders and the group cohomology of their symmetry group," Science \textbf{338}, 1604 (2012); arXiv:1106.4772v6.


\bibitem{Li} H. Li and F.~D.~M. Haldane, ``Entanglement Spectrum as a Generalization of Entanglement Entropy: Identification of Topological Order in Non-Abelian Fractional Quantum Hall Effect States," Phys. Rev. Lett. {\bf 101}, 010504 (2008).


\bibitem{MERA} G.~Vidal, ``Class of Quantum Many-Body States That Can Be Efficiently Simulated," Phys. Rev. Lett. {\bf 101}, 110501 (2008).


\bibitem{Swingle:2012wq}
  B.~Swingle,  ``Constructing holographic spacetimes using entanglement renormalization,''
  arXiv:1209.3304 [hep-th].
	

	
\bibitem{Bennett:1992tv}
  C.~H.~Bennett, G.~Brassard, C.~Crepeau, R.~Jozsa, A.~Peres and W.~K.~Wootters,
  ``Teleporting an unknown quantum state via dual classical and Einstein-Podolsky-Rosen channels,''
  Phys.\ Rev.\ Lett.\  {\bf 70}, 1895 (1993).

\bibitem{Bennett:1992dc} C.~H.~Bennet, S. Wiesner,   "Communication via one- and two-particle operators on Einstein-Podolsky-Rosen states," Phys.\ Rev.\ Lett.\  {\bf 69}, 2881 (1992)


\bibitem{Laiho:2000mc}
  R.~Laiho, S.~N.~Molotkov and S.~S.~Nazin,
  ``Teleportation of the relativistic quantum field,''
  Phys.\ Lett.\ A {\bf 275}, 36 (2000)
  [quant-ph/0005067].




\bibitem{Hotta:2008uk}
  M.~Hotta,
  ``Quantum measurement information as a key to energy extraction from local vacuums,''
  Phys.\ Rev.\ D {\bf 78}, 045006 (2008)
  [arXiv:0803.2272 [physics.gen-ph]].

\bibitem{Hotta:2009fz}
  M.~Hotta,
  ``Controlled Hawking Process by Quantum Energy Teleportation,''
  Phys.\ Rev.\ D {\bf 81}, 044025 (2010)
  [arXiv:0907.1378 [gr-qc]].

\bibitem{Hotta:2009ppa}
  M.~Hotta,
  ``Quantum Energy Teleportation with Electromagnetic Field: Discrete vs. Continuous Variables,''
  arXiv:0908.2674 [quant-ph].


\bibitem{Hotta:2011xj}
  M.~Hotta,
  ``Quantum Energy Teleportation: An Introductory Review,''
  arXiv:1101.3954 [quant-ph].

\bibitem{Giataganas:2016bml}
  D.~Giataganas, F.~L.~Lin and P.~H.~Liu,
  ``Towards Holographic Quantum Energy Teleportation,''
  Phys.\ Rev.\ D {\bf 94}, no. 12, 126013 (2016)
  [arXiv:1608.06523 [hep-th]].

\bibitem{Miyaji:2015yva}
  M.~Miyaji and T.~Takayanagi,
  ``Surface/State Correspondence as a Generalized Holography,''
  PTEP {\bf 2015}, no. 7, 073B03 (2015)
  [arXiv:1503.03542 [hep-th]].


\bibitem{Miyaji:2015fia}
  M.~Miyaji, T.~Numasawa, N.~Shiba, T.~Takayanagi and K.~Watanabe,
 ``Continuous Multiscale Entanglement Renormalization Ansatz as Holographic Surface-State Correspondence,''
  Phys.\ Rev.\ Lett.\  {\bf 115}, no. 17, 171602 (2015)
  [arXiv:1506.01353 [hep-th]].

\bibitem{opeblock}
  B.~Czech, L.~Lamprou, S.~McCandlish, B.~Mosk and J.~Sully,
  ``A Stereoscopic Look into the Bulk,''
  arXiv:1604.03110 [hep-th].


\bibitem{shadows}
  D.~Simmons-Duffin,
  ``Projectors, Shadows, and Conformal Blocks,''
  JHEP {\bf 1404}, 146 (2014)
  [arXiv:1204.3894 [hep-th]].



\bibitem{Calabrese:2006rx}
  P.~Calabrese and J.~L.~Cardy,
  ``Time-dependence of correlation functions following a quantum quench,''
  Phys.\ Rev.\ Lett.\  {\bf 96}, 136801 (2006)
  [cond-mat/0601225].

\bibitem{Hartman:2015lfa}
  T.~Hartman, S.~Jain and S.~Kundu,
  ``Causality Constraints in Conformal Field Theory,''
  JHEP {\bf 1605}, 099 (2016)
  [arXiv:1509.00014 [hep-th]].




\end{thebibliography}
\end{document}